# SUBSTANTIAL UPGRADES TO TEVATRON LUMINOSITY

Mike Church, Fermilab, PO Box 500 Batavia, IL, 60510

Over the next 6 years the CDF and D0 collaborations have the goal of collecting 15 fb$^{-1}$ of data in collider Run II. This will require a number of ambitious upgrades to the Fermilab accelerator complex, some of which have been completed, some of which are currently being commissioned, and some of which are under development but have not yet been installed. This paper describes the major accelerator upgrades required to reach this Run II luminosity goal.

# 1  INTRODUCTION

In Run Ib, from 1993 to 1996, the Fermilab accelerator complex delivered a total of .14 fb$^{-1}$ of integrated luminosity to the CDF and D0 detectors at a center-of-mass energy of 1.8 TeV.  During this run the discovery of the top quark was announced.  The Tevatron operated in a mode colliding 6x6 proton and $\bar{p}$ bunches.  The greatest limitation to luminosity was, and is expected to remain, the $\bar{p}$ production rate. $\bar{p}$ production rates up to 7.2x10$^{10}$ $\bar{p}$/hour were attained in Run Ib.[1]

After 1996 the accelerator complex was shut down for the construction of two new rings.  The Main Injector (MI) replaced the original Main Ring for accelerating particles from 8 GeV to 150 GeV, and the Recycler Ring (RR) was constructed in the MI tunnel.[2]  The RR is an 8 GeV permanent magnet storage ring to be used as a post-Accumulator storage ring for $\bar{p}$'s and to store $\bar{p}$'s that remain unused at the end of a collider store.

In 2000 a Run II "engineering run" took place, which established 36x36 operation at a center-of-mass energy of 1.96 TeV and delivered luminosity to the CDF detector for commissioning.  Run II proper began in March of 2001, and luminosity is currently being delivered to the CDF and D0 detectors.

Table 1 shows the machine parameters for Run Ib and Run II.  Run II is divided into two distinct stages:  Run IIa will collide 36x36 proton and $\bar{p}$ bunches, and Run IIb will collide 140x103 proton and $\bar{p}$ bunches.  The increase in number of bunches will keep the total number of interactions/crossing at a tolerable level for the collider detectors as the luminosity increases.

The planned luminosity upgrades are the following:  1) increase Booster integrated intensity; 2) increase MI intensity on $\bar{p}$ production cycles; 3) complete commissioning of the RR; 4) increase the $\bar{p}$ yield from the $\bar{p}$ production target; 5) upgrade Debuncher stochastic cooling; 6) upgrade Accumulator stochastic cooling; 7) increase the Tevatron energy; 8) commission the Tevatron Electron Lens (TEL); and 9) implement crossing angles at the interaction regions (IR's).

Table 1: Run Ib and Run II parameters

| RUN | Ib (6x6) | IIa (36x36) | IIb (140x103) | units |
|---|---|---|---|---|
| protons/bunch | $2.3 \times 10^{11}$ | $2.7 \times 10^{11}$ | $2.7 \times 10^{11}$ | |
| antiprotons/bunch | $5.5 \times 10^{10}$ | $3.0 \times 10^{10}$ | $10.0 \times 10^{10}$ | |
| total antiprotons | $.33 \times 10^{12}$ | $1.1 \times 10^{12}$ | $11.0 \times 10^{12}$ | |
| antiproton production rate | $6.0 \times 10^{10}$ | $10. \times 10^{10}$ | $52. \times 10^{10}$ | hr$^{-1}$ |
| 95% normalized proton emittance | $23\pi$ | $20\pi$ | $20\pi$ | mm-mrad |
| 95% normalized antiproton emittance | $13\pi$ | $15\pi$ | $15\pi$ | mm-mrad |
| $\beta^*$ | 35 | 35 | 35 | cm |
| energy | 900 | 980 | 980 | GeV |
| antiproton bunches | 6 | 36 | 103 | |
| bunch length (rms) | .60 | .37 | .37 | m |
| crossing angle at IR | 0 | 0 | 136 | μrad |
| typical luminosity | $.16 \times 10^{32}$ | $.86 \times 10^{32}$ | $5.2 \times 10^{32}$ | cm$^{-2}$sec$^{-1}$ |
| integrated luminosity | 3.2 | 17.3 | 105 | pb$^{-1}$/week |
| bunch spacing | ~3500 | 396 | 132 | nsec |
| interactions/crossing | 2.5 | 2.3 | 4.8 | |

## 2 OVERVIEW OF OPERATIONS

Figure 1 shows the layout of the Fermilab Accelerator complex relevant to collider operation. The Booster Ring accelerates protons from 400 MeV to 8 GeV and injects them into the MI. The MI does one of three things with these protons: accelerates them from 8 GeV to 150 GeV and injects them into the Tevatron for collider operation; accelerates them from 8 Gev to 120 GeV and extracts them to the $\bar{p}$ production target for $\bar{p}$ production; or delivers them at 8 GeV to the RR and the Antiproton Source for commissioning studies. 8 GeV $\bar{p}$ 's produced at the $\bar{p}$ production target are collected in the Debuncher Ring, debunched, stochastically cooled, and then transferred to the Accumulator Ring, where a large $\bar{p}$ "stack" is accumulated and stochastically cooled. When a suitably large stack is accumulated, these are transferred to the RR for further storage at 8 GeV. At the start of a store, the $\bar{p}$ 's are transferred from the RR to the MI, accelerated to 150 GeV and then injected into the Tevatron. The Tevatron accelerates protons and $\bar{p}$'s simultaneously to 980 GeV. At the end of a store the protons are scraped away with beam collimators, and the $\bar{p}$'s are decelerated down to 150 GeV, injected into the MI, decelerated down to 8 GeV, and injected into the RR where they are recooled for use in a later store.

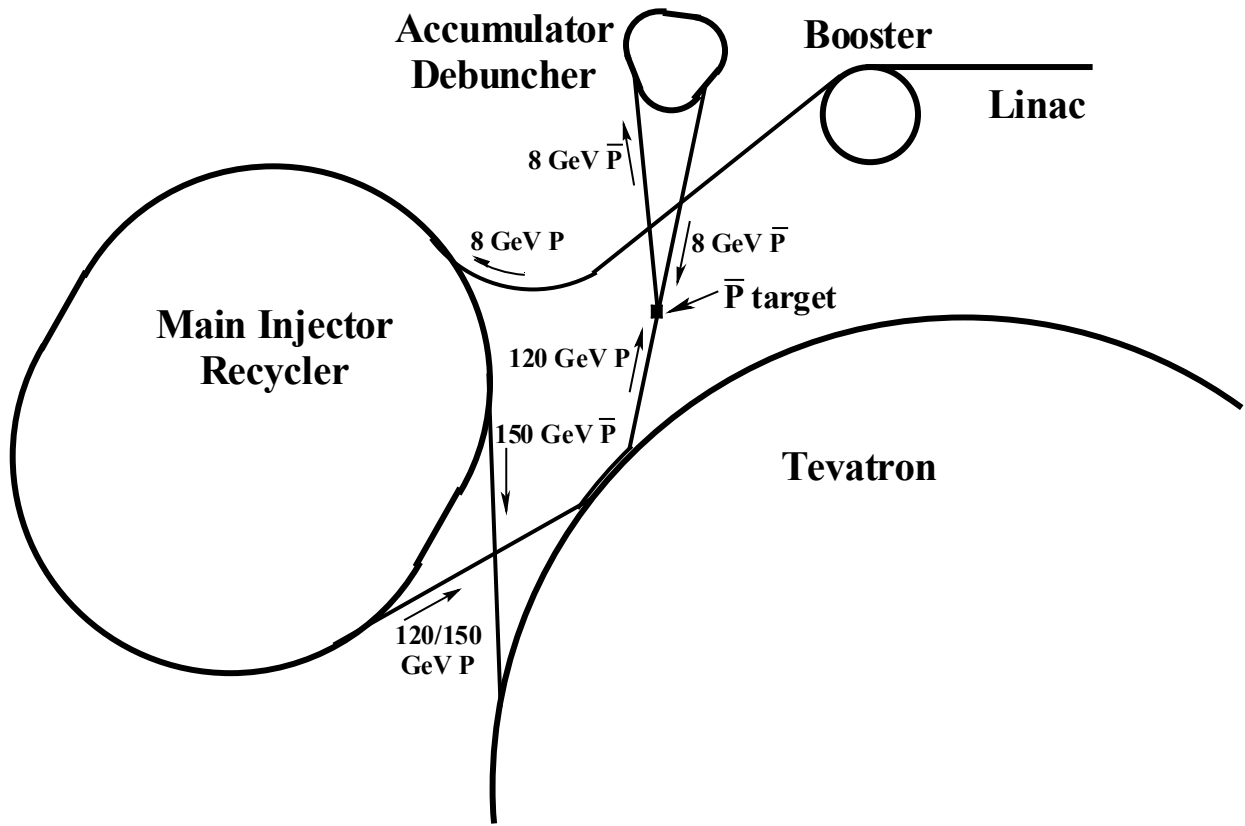

Figure 1: Fermilab Accelerator complex for collider operation

## 3 BOOSTER UPGRADES

Currently Booster intensity is limited by integrated radiation losses due to excessive beam losses.[3] For Run IIa the required intensity is $5.2 \times 10^{12}$ protons/pulse @ 1 pulse/1.5 seconds, and for Run IIb the required intensity is $5.2 \times 10^{12}$ protons/pulse @ 2 pulses/1.5 seconds. The current record is $4.3 \times 10^{12}$ protons/pulse @ 1 pulse/1.5 seconds with acceptable radiation losses. In progress are the following upgrades to reduce losses: 1) extraction aperture increase; 2) installation of ramped corrector magnets; 3) increased radiation shielding; 4) notching the beam to reduce septum magnet losses; and 5) installation of beam collimators to localize losses in "safe" areas.

## 4 MAIN INJECTOR UPGRADE

The MI is currently operating within 20% of its design intensity of $5.0 \times 10^{12}$ protons/pulse on $\bar{p}$ production cycles. This intensity will be increased by a factor of ~1.8 by implementing "slip stacking". In this scheme, two batches of 84 proton bunches are injected from the Booster on each MI acceleration cycle, each with a small momentum offset. These two batches are controlled by independent RF systems. As the batches cog, or "slip" past one another in the ring, the RF frequencies are brought close to each other, and then the two batches are recaptured in a single RF system. Low intensity slip stacking has already been demonstrated in the old Main Ring, however high intensity slip stacking requires beam loading compensation on the MI RF cavities, which is currently in progress.

## 5 ANTIPROTON SOURCE UPGRADES

During Run Ib the typical $\bar{p}$ yield into the Debuncher from the $\bar{p}$ production target was $21 \times 10^{-6}$ $\bar{p}$/proton on target. The goal is to increase this to $34 \times 10^{-6}$ for Run II by increasing the lithium lens gradient, and by increasing the downstream aperture between the Debuncher and the $\bar{p}$ production target. $\bar{p}$'s produced at the production target are focused into a beamline by a pulsed lithium lens.[4] Thermal stresses in the lithium lens body, induced by the ohmic losses from the high current pulse, have prevented the lens from being operated reliably above a gradient of ~760 T/m. Mechanical modifications to the lens will permit reliable operation at 900 T/m. In addition, the downstream aperture will be increased from ~19π-mm-mrad to ~32π-mm-mrad through beampipe modifications, a larger aperture septum magnet, and improved beamline and Debuncher steering.

To support collider operation the $\bar{p}$ phase space density must be increased by a factor of $10^8$ from the production target to the Accumulator stack. Since the cooling rate of stochastic cooling systems scales with the cooling bandwidth squared, the most straightforward way to increase cooling rate is to increase the cooling system bandwidth. The 2-4 GHz bandwidth Debuncher stochastic cooling system has been replaced with a 4-8 GHz system. The broadband 2-4 GHz planar loop structures have been replaced by 4 narrow band slotted waveguide structures covering the 4-8 GHz band.[5] This new cooling system is designed to cool $80 \times 10^{10}$ $\bar{p}$/hr – which is adequate for Run IIb.

The Accumulator 1-2 GHz stacktail cooling system has been replaced with a 2-4 GHz system. In order to accommodate this upgrade, the Accumulator lattice was modified – the slip factor, $\eta$, was decreased from .023 to .012. This upgrade is adequate for Run IIa, but a further increase in cooling bandwidth to 4-8 GHz will be required for Run IIb.

## 6 RECYCLER RING STATUS

The RR is currently being commissioned with protons. Major difficulties encountered during this commissioning stage have been: 1) stray magnetic fields from the MI; 2) magnet, beampipe, and BPM misalignment; and 3) field errors in the magnets. The lattice is now well-understood, although magnet misalignment is requiring the installation of additional dipole correction magnets in the ring. The current transverse aperture is ~$10\pi$-mm-mrad, whereas the design aperture is $40\pi$-mm-mrad, and >$20\pi$-mm-mrad is probably necessary for Run IIa. Stray magnetic fields are being overcome with the addition of more magnetic shielding – beam lifetimes of >5 hours have been measured with the MI in a ramping state. The stochastic cooling tanks have been installed and beam cooling observed with $\bar{p}$'s.

In Run IIb the high $\bar{p}$ intensities will require high energy electron cooling in the RR -- 4.3 MeV electrons @ .5A. A test using an older pelletron[6] gave a stable .7A with a 1.5 MeV beam. The current electron cooling system is being built and commissioned outside of the RR and will be installed in the RR in 2003 for Run IIb.

## 7 TEVATRON UPGRADES

The Tevatron energy has recently been increased from 900 GeV to 980 GeV. This was accomplished by implementing several modifications over the last 8 years:[7]

1) Magnets with low quench currents have been selectively removed from the ring or moved to colder locations in the ring,

2) Twenty-four He cold compressors have been installed which allow Tevatron operation with 2-$\phi$ He temperatures between 3.9 and 3.6 ºK.

3) Recoolers – spool pieces with 1-$\phi$ to 2-$\phi$ heat exchangers – have been installed to provide more cold locations in the ring.

With 140x103 operation during Run IIb the bunch spacing will be reduced to 132 nsec. Without crossing angles at the IR's there will be 3 beam crossings at each low beta region which will give an unacceptably large beam-beam tune shift, therefore crossing angles at the IR's are required. The 3 additional electrostatic separators which will be required are currently being built.

An electron beam will be used to compensate for the beam-beam tune shift spread in Run IIb. This device, called the Tevatron Electron Lens[8], has been installed in the Tevatron at the F48 straight section and commissioning has begun. A 10KV, 1A electron beam is run coaxially with the $\bar{p}$ beam for 2 meters. The focusing effect of the electrons compensates for the tune shift caused by protons (both long range and short range). This will permit either higher proton intensities or a reduction of the crossing angles at the IR's in Run IIb.

## 8 LUMINOSITY PREDICTIONS

The upgrades discussed above can be included in a numerical calculation of integrated luminosity in the Tevatron under varying assumptions: beam transfer efficiencies, machine downtime, emittance growth rates, time between stores…. Figure 2 shows the weekly integrated luminosity vs $\bar{p}$ stacking rates under a variety of operational scenarios. To meet the Run IIb goal requires a weekly integrated luminosity of about 115 pb$^{-1}$/week.

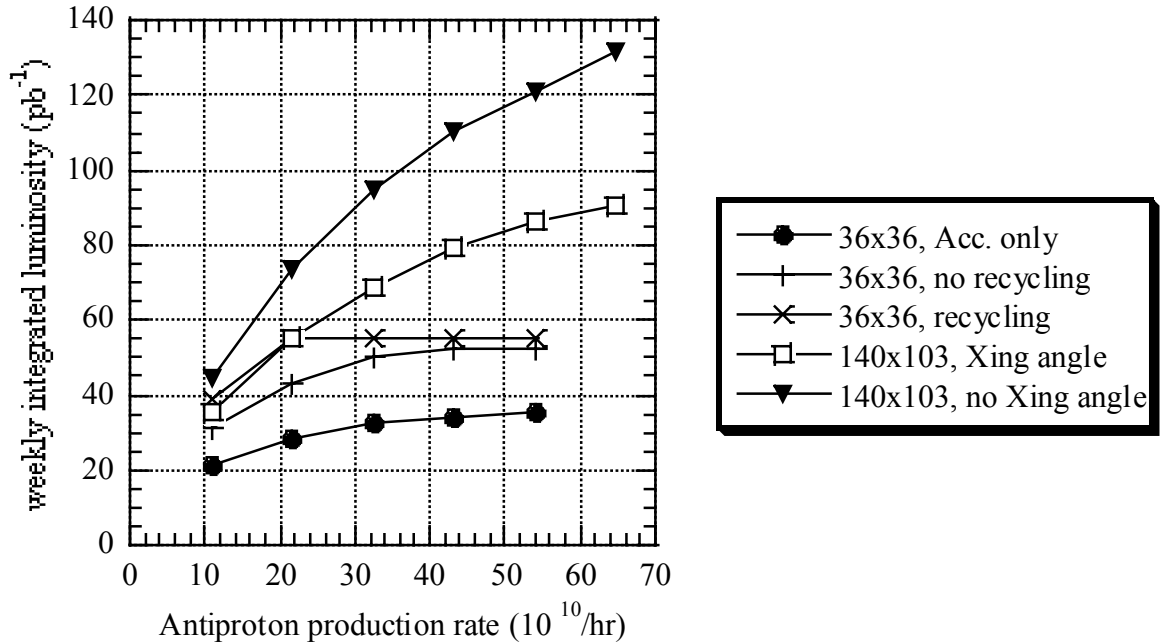

Figure 2: Weekly integrated luminosity vs. stacking for Run II

## 9 REFERENCES

1) V Bharadwaj, et al, "Fermilab Collider Run 1b Accelerator Performance", FERMILAB-TM-1970 (1996)

2) G Jackson, "The Fermilab Recycler Ring Technical Design Report", FERMILAB-TM-1991 (1996)

3) R Webber, "Operational Experience with Beam Loss, Shielding, and Residual Radiation in the Fermilab Proton Source", FERMILAB-CONF-00-193 (2000)

4) J Marriner, M Church, "The Antiproton Sources: Design and Operation", Ann. Rev. Nucl. Part Sci., V43, p253-95, Annual Review Inc., Palo Alto, Ca (1993)

5) D McGinnis, "Slotted Waveguide Slow Wave Stochastic Cooling Arrays, Fermilab Pbar Note 626, (1999)

6) S Nagaitsev, et al, "Successful MeV-range Electron Beam Recirculation", Proceedings of the 6th European Particle Accelerator Conference, Stockholm, Sweden (1998)

7) J Theilacker, "Cryogenic Testing and Analysis Associated with Tevatron Lower Temperature Operation", FERMILAB-CONF-96-289, (1996)

8) D Shatilov,V Shiltsev, "Simulations of the Tevatron Beam Dynamics with Beam-beam Compensation" FERMILAB-TM-2124 (2000)